\documentclass[12pt
]
{iopart}

\usepackage{graphicx,color}
\usepackage{setstack}
\usepackage{amssymb}
\graphicspath{{Diagrams/}}

\definecolor{ngreen}{rgb}{0.2,0.6,0.2}
\definecolor{nred}{rgb}{0.6,0.2,0.2}

\newcommand{\beq}{\begin{equation}}
\newcommand{\eeq}{\end{equation}}
\newcommand{\bqa}{\begin{eqnarray}}
\newcommand{\eqa}{\end{eqnarray}}

\newcommand{\erf}[1]{Eq.~(\ref{#1})}

\newcommand{\erfand}[2]{Eqs.~(\ref{#1}) and (\ref{#2})}

\newcommand{\cu}[1]{\left\{ {#1} \right\}}

\newcommand{\an}[1]{\left\langle{#1}\right\rangle}

\newcommand{\ket}[1]{\ensuremath{\left| #1 \right\rangle}}
\newcommand{\bra}[1]{\ensuremath{\left\langle #1 \right|}}

\newcommand{\nn}{\nonumber}

\begin{document}

\title{Measuring measurement--disturbance relationships with weak values}
\author{A P Lund and H M Wiseman}
\address{Centre for Quantum Computer Technology, Centre for Quantum
Dynamics, Griffith University, Nathan Queensland 4111, Australia}
\eads{\mailto{A.Lund@Griffith.edu.au} and \mailto{H.Wiseman@Griffith.edu.au}}

\begin{abstract}
Using formal definitions for measurement precision $\epsilon$ and
disturbance (measurement backaction) $\eta$, Ozawa [{Phys. Rev. A}
{\bf 67}, 042105 (2003)] has shown that Heisenberg's claimed relation
between these quantities is false in general. Here we show that the
quantities introduced by Ozawa can be determined experimentally, using
no prior knowledge of the measurement under investigation --- both
quantities correspond to the root-mean-squared difference given by a weak-valued
probability distribution. We propose a simple three-qubit experiment
which would illustrate the failure of Heisenberg's
measurement--disturbance relation, and the validity of an alternative
relation proposed by Ozawa.
\end{abstract}

\pacs{03.65.Ta,03.67.Ac}

\submitto{\NJP}

\maketitle

\section{Introduction}

The Heisenberg-Robertson uncertainty relation \cite{Hei27,Rob29}
constrains the standard deviations (SDs) of two arbitrary
non-commuting observables $A$ and $B$~\cite{fn1}: 
\beq  \label{HRUR}
\sigma(A)\sigma(B) \geq 
\frac{1}{2} \left| \bra{\psi} [A,B] \ket{\psi} \right|.
\eeq
The foremost example is for canonically conjugate observables,
$\sigma(q)\sigma(p) \geq \hbar/2$, as considered by Heisenberg
\cite{Hei27}. The uncertainty relation of \erf{HRUR} is now
uncontroversial and is well verified experimentally
\cite{experiment:shull, experiment:nairz}. However, this was not the
only relation introduced by Heisenberg in Ref.~\cite{Hei27}; the first
relation discussed there involved quite different quantities:
$\epsilon$, the {\em precision} with which a quantity is measured, and
$\eta$, the amount of {\em disturbance} (Heisenberg called this the
{\em discontinuous change}) in some other quantity. The status of this
measurement--disturbance relation (MDR) has been a matter of
considerable debate \cite{ScuEngWal91,StoTanColWal94,WisHar95}.

Heisenberg at first considered only position measurement and momentum
disturbance, and postulated the relation $\epsilon(q)\eta(p) \gtrsim
h$, giving a heuristic derivation from a description of Compton
scattering \cite{Hei27}. Later \cite{Hei30}, he rigorously derived the
MDR
\beq \label{HeiMDR}
\epsilon(q)\eta(p) \geq \hbar/2,
\eeq
but only for a very special case~\cite{Wis98}, {\em viz.} where
initially the particle was in a momentum eigenstate (and thus had
completely undefined position) and the measurement apparatus performs
a quantum non-demolition measurement. In this case the measurement
precision $\epsilon(q)$ can be identified with the post-measurement
position SD $\sigma'(q)$, and the momentum disturbance can be
quantified by the post-measurement momentum SD $\sigma'(p)$, so that
\erf{HeiMDR} follows from \erf{HRUR}. For other types of initial
states or measurement apparatus, it is not at all obvious how these
quantities should be defined \cite{WisHar95,Wis98}, or even that the
analogue of \erf{HeiMDR} should be expected to hold~\cite{Ozawa02}.

It was argued by Scully and co-workers \cite{ScuEngWal91} that, in the
context of a twin-slit interferometer, one can perform a position
measurement with sufficient precision to determine which way the
particle goes, without disturbing its momentum at all. In such
experiments, $\eta(p)$ is indeed zero, if one defines this quantity as
the root-mean-squared (RMS) difference of the so-called weak-valued
probability distribution for momentum disturbance \cite{Wis03}. This
is a distribution that can be, and indeed recently has been
\cite{Mir07}, directly observed experimentally. It thus seems that the
MDR of \erf{HeiMDR} is not valid in general.

This conclusion, that a MDR of the Heisenberg form was not universally
valid, was independently arrived at by Ozawa~\cite{Ozawa03}. But Ozawa
went further, and proposed a new, universally valid,
MDR~\cite{Ozawa03}. Moreover, Ozawa's MDR applies for any pair of
observables $A$ (which is measured) and $B$ (which is disturbed).
However, unlike in Ref.~\cite{Wis03}, the quantities of Ozawa's MDR
were defined purely theoretically. That is, no prescription was given
in Ref.~\cite{Ozawa03} for how these quantities could be
experimentally determined given (i) a black-box apparatus which
performs some sort of measurement of $A$, on a system prepared in some
fixed but {\em unknown mixed state} $\rho$. In a later
paper~\cite{Ozawa04}, Ozawa gave a method for determining the
measurement precision $\epsilon$, however this method requires the
system to be prepared in a {\em known pure state} $\ket{\psi}$.

In this paper we unify the approach of Ref.~\cite{Wis03} with that of
Ref.~\cite{Ozawa03}. First we show that Ozawa's disturbance quantifier
$\eta(B)$ is exactly the RMS difference given by a weak-valued
probability distribution for disturbance defined in Ref.~\cite{Wis03}.
Second, we show that Ozawa's precision quantifier $\epsilon(A)$ equals
the RMS difference given by another, analogously defined, weak-valued
probability distribution. The remaining quantities in Ozawa's relation
are simply the SDs of observables for the initial state $\rho$. Thus,
Ozawa's MDR, for $B$-disturbance caused by an $A$-measurement, could
be experimentally tested by an experimenter without knowledge of the
initial system state, the initial meter state, or of the interaction.
Finally, we propose a simple three-qubit experiment which could
demonstrate ``interesting'' cases where Ozawa's MDR would be
validated, but a Heisenberg-form MDR would be violated.

\section{Ozawa's MDR}
\begin{figure}
\centerline{\includegraphics[width=9cm]{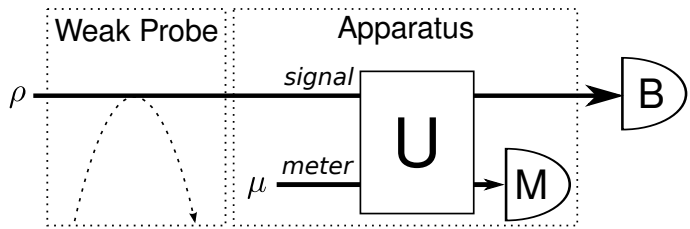}}
\caption{\label{schematic}Schematic of the procedure using weak
measurements to extract the measurement disturbance and precision
quantities. An initial weak measurement shown in the left hand box is
used to gather information about the signal prior to its measurement
by the apparatus. This apparatus can be described without loss of
generality as the preparation of a meter state $\mu$, the interaction
of the signal and meter by a unitary $U$, and the read-out of some
meter observable $M$. Measurement results from the weak probe, and
from the read-out of the meter or a final (third) measurement of the
signal, are used to construct the measurement disturbance and
precision quantities.}
\end{figure}
Using the measurement model described in Fig.~\ref{schematic}, Ozawa's
precision and disturbance quantities are defined as~\cite{Ozawa03}:
\bqa
\label{precision}
\epsilon(A) &=& \left< \left( U^\dagger(I \otimes M)U
 - A \otimes I \right)^2 \right>^{1/2}, \\
\label{disturbance}
\eta(B) &=& \left< \left( U^\dagger(B \otimes I)U
 - B \otimes I \right)^2 \right>^{1/2}.
\eqa
where the average is taken over the composite input signal and meter
state. Here $U$ is the unitary evolution operator describing the
complete interaction between signal and meter, which is intended to
imprint information about the signal observable $A$ onto the meter
observable $M$ (which is read-out), while $B$ is a signal observable
that may (or may not) be disturbed as a result of the measurement of
$A$. The quantity $\epsilon(A)$ is the RMS difference between the
initial value of $A$ and the final value of $M$, which is the obvious
definition for {\em measurement precision} ~\cite{Ozawa03}. Similarly,
$\eta(B)$ is the RMS difference between the initial and final values
of $B$ --- the {\em disturbance} of $B$ caused by measurement
backaction.

One might expect that these quantities would satisfy
a  Heisenberg-form  MDR 
\beq \label{falseMDR}
H \equiv \epsilon(A) \eta(B) \geq C(A,B) ,
\eeq
where $C(A,B)$ derives from the commutator in \erf{HRUR},
\beq \label{unc-bound}
C(A,B) \equiv \frac{1}{2}\Big| \Tr \big\{\rho [A,B] \big\} \Big|.
\eeq
Heisenberg's original relation \erf{HeiMDR} would then be a special
case. However, as shown by Ozawa, this relation is false in general.
Ozawa then suggested replacing the simple Heisenberg product $H$ in
\erf{falseMDR} by a three-term expression involving not only
$\epsilon(A)$ and $\eta(B)$, but also the SDs appearing in \erf{HRUR}:
\beq \label{OzawaO}
O \equiv \epsilon(A) \eta(B) + \epsilon(A) \sigma(B) + \sigma(A) \eta(B) .
\eeq
This enabled him to derive a universally valid MDR~\cite{Ozawa03}:
\beq \label{OzawaMDR}
O \geq C(A,B)
\end{equation}
Note that $\sigma(A)$ and $\sigma(B)$ are evaluated from measurements
upon the initial signal state $\rho$.

\section{Ozawa's MDR and Weak Values}
Ozawa's precision and disturbance quantities both involve observables
at two different times --- before and after the signal--probe
interaction. Classically, one could determine such a quantity simply
by measuring the relevant observables at the two times. Quantum
mechanically, the problem is of course that the first measurement
would disturb the system, so rendering the result of the second
measurement irrelevant. This problem can be overcome by using weak
measurements~\cite{AhaAlbVai88}. Such measurements may give
vanishingly small disturbances of the underlying system, at the cost
of having a very noisy result, and so requiring a very large ensemble
to obtain an accurate average. The average of a weak measurement
result is particularly interesting when performed on an ensemble
post-selected on a later, usually strong, measurement. Such averages
are called weak-values~\cite{AhaAlbVai88}, and have been used to
analyse a great variety of quantum phenomena~\cite{Mir07,RitStoHul91,Ste95,Wis02a,RohAha02,Brun03,Sol04,Pry05,Wis07,HosKwi08}.
We will now show how these weak measurements can be utilised to
extract the quantities in the universally valid MDR.

First, consider the $\eta(B)$ quantity. For simplicity, we will take
the eigenvalue spectrum $\cu{b}$ of $B$ to be discrete; the
generalisation to a continuous spectrum is not difficult to construct
theoretically, but in an experiment discretization would be
necessary~\cite{Mir07}. We denote the projector onto the eigenspace
associated with eigenvalue $b$ as $\Pi(b)$. These projectors are
Hermitian and hence are valid observables. The expectation value $\Tr
\left[\Pi(b) \rho \right]$ is the probability that the system would be
found to have $B=b$, via a strong measurement of the observable
$\Pi(b)$. The same number, between zero and one, would be obtained by
averaging a weak measurement of $\Pi(b)$.

The equivalence between averages of weak and strong measurements of
$\Pi(b)$ ceases when one considers post-selected ensembles. The
general expression for a post-selected weak-value, allowing for a
mixed initial state, arbitrary evolution, and an arbitrary final
measurement was given in Ref.~\cite{Wis02a}. In this case the
evolution in question is the interaction between signal and meter,
and, since we are interested in the disturbance to $B$, the final
measurement we wish to make is one of the observable $B$. The weak
value of the projection observable $\Pi(b)$, post-selected on
achieving a final result $b_f$, is
\beq \label{weak-value}
\ _{b_f} \left< \Pi(b_i)_{\rm weak} \right>_\rho = \mathrm{Re}
\frac{\Tr \left[ \Pi(b_f) U \Pi(b_i) (\rho\otimes\mu) U^\dagger \right]}
 {\Tr \left[\Pi(b_f) U (\rho\otimes\mu) U^\dagger \right]}.
\eeq
Here $\mu$ is the initial meter state, and $\Pi(b)$ is to be
understood as $\Pi(b)\otimes I$. We have used subscripts $i$ and $f$
to explicitly indicate the values representing quantities before ({\em
i}nitial) and after ({\em f}inal) the unitary evolution. The
expression in \erf{weak-value} can be interpreted as the weak-valued
probability of $B$ initially taking eigenvalue $b_i$, conditional on
it finally taking eigenvalue $b_f$. Hence we will write this
expression as $P_\mathrm{\rm wv}(b_i | b_f)$. Using this, it is
possible to define a weak valued {\em joint} probability
distribution~\cite{Wis03}
\bqa
P_\mathrm{\rm wv}(b_i, b_f) &=& P_\mathrm{\rm wv}(b_i | b_f) P(b_f) \nonumber \\
&=& \mathrm{Re} \left\{
 \Tr \left[ \Pi(b_f) U \Pi(b_i) (\rho\otimes\mu) U^\dagger \right]
 \right\} \nonumber \\
&=& \mathrm{Re} \left< U^\dagger \Pi(b_f) U \Pi(b_i) \right>.
\eqa
Finally, we define a weak-valued probability distribution for a change
$\delta b$ in the value of observable $B$ as
\bqa
P_{\rm wv}(\delta b) &=& \sum_{b_i} P_{\rm wv}(b_i, b_f=b_i + \delta b) \label{using} \\
\label{wvpB}
&=& \sum_{b} {\rm Re} \left< U^\dagger \, \Pi(b+\delta b) \, U \, \Pi(b)  \right>. 
\eqa

We will now show that the RMS difference given by this weak-valued
probability distribution for the change in $B$ is identical to Ozawa's
disturbance quantity $\eta(B)$. Making the signal and meter Hilbert
spaces explicit, the mean squared change in $B$ is
\begin{eqnarray}
 && \sum_{\delta b} (\delta b)^2 P_{\rm wv}(\delta b) \label{prec-delta} \\
&=& \sum_{b,b^\prime} (b^\prime - b)^2
    {\rm Re}  \left< U^\dagger \left[\Pi(b^\prime) \otimes I \right]
    U \left[\Pi(b) \otimes I \right]  \right> \nonumber \\
&=& \left< \left(U^\dagger \left[ B \otimes I \right] U -
    \left[ B \otimes I \right]\right)^2   \right>  \label{finaljust}
\end{eqnarray}
where we have used the Hermitian operator identity $f(B) = \sum_b f(b)
\Pi(b)$ and written out the real part using the sum of complex
conjugate pairs in the last step. The final equation (\ref{finaljust})
is just the square of Eq.~(\ref{disturbance}).

The measurement precision quantity $\epsilon(A)$ can be measured
similarly. For each eigenvalue $a$, one performs a weak measurement of
$\Pi(a)$ on the initial signal before it enters the apparatus, and
then a strong measurement (read-out) of the final meter observable
$M$, which contains the information about $A$. As above, one can then
construct the weak-valued probability distribution for the difference
between the initial value of $A$ and the final value of $M$. Since the
meter is meant to measure $A$, we can assume, following
Ref.~\cite{Ozawa03}, that the spectrum of $M$ coincides with that of
$A$. Then this weak-valued probability distribution evaluates to the
following (note the difference in the ordering of tensor products in
the two terms):
\begin{equation}
P_{\rm wv}(\delta a) = \sum_a {\rm Re} \left< U^\dagger
\left[I \otimes \Pi(a+\delta a)\right] U \left[\Pi(a) \otimes I
  \right] \right> .\label{wvpA}
\end{equation}
Following a similar computation to the disturbance case,
\begin{equation}
\label{weak-prec}
\sum_{\delta a} (\delta a)^2 P_{\rm wv}(\delta a) = 
\left< \left(U^\dagger \left[ I \otimes M \right] U 
     - \left[A \otimes I\right]\right)^2   \right>
\end{equation}
is square of \erf{precision}. Thus, both of Ozawa's quantities can be
measured using a weak measurement immediately before the measurement
under examination is applied, without any assumptions about $\rho$,
$\mu$, or $U$. See Fig.~\ref{schematic}.

A notable property of our experimental method of determining
of the measurement precision and disturbance quantities, via
weak-valued probability distributions, is that they coincide with how
a classical physicist might perform such a determination
~\cite{Gar04}. Of course classically one could just measure the
initial value of the observable and the final value and then after
gathering statistics find the RMS difference. However, this method
does not directly transfer to quantum mechanics because strong
projective measurement will disturb the statistics gathered. Using
weak measurements gives a valid quantum mechanical procedure, which is
also applicable to classical systems, in which case it would give
exactly the same answer as would be obtained using strong
measurements. This is another distinction from the experimental method
proposed by Ozawa \cite{Ozawa04}.

\section{Qubit example}
We will now construct an example using qubits. As usual
\cite{NieChu00}, we write the Pauli matrices as $X$, $Y$ and $Z$, and
the states $\ket{0}$ and $\ket{1}$ denote $Z$ eigenstates with
eigenvalues of $1$ and $-1$ respectively.

\begin{figure}
\centerline{\includegraphics[width=12cm]{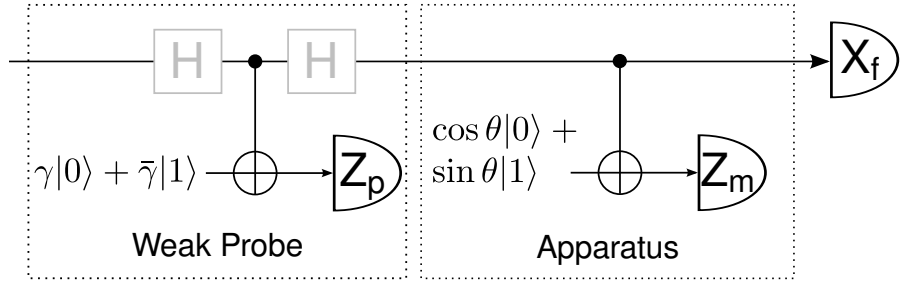}}
\caption{\label{qubit}A qubit-based implementation of the device shown
in Fig.~\ref{schematic}. When measuring the disturbance, the probe
measurement is conjugated by Hadamard operations (shown in grey) to
perform a weak $X$ measurement. For the precision they are absent and
the probe performs a weak $Z$ measurement.}
\end{figure}

We take the observable to be measured (the operator $A$) to be $Z$.
The measurement apparatus is constructed by choosing the measurement
interaction (the unitary $U$) to be the $CNOT$ operation
\cite{NieChu00}, a pure initial meter state $\mu =
\ket{\theta}\bra{\theta}$, where $\ket{\theta} = \cos\theta \ket{0} +
\sin\theta \ket{1}$, and the meter measurement observable (the
operator $M$) to be $Z$; see Fig.~\ref{qubit}. The strength of this
measurement can be quantified as $\cos2\theta$, varying from a full
strength $Z$-measurement at $\theta=0$ and no measurement at
$\theta=\pi/4$~\cite{Pry05}. To test the two MDRs (\ref{falseMDR}) and
(\ref{OzawaMDR}) we consider the disturbance in the signal observable
$X$ (i.e. this is the operator $B$). This choice allows for the
maximum value of the $C(A,B)$ of \erf{unc-bound}, which is the
lower-bound appearing in both MDRs. This maximum is achieved when the
input signal state $\rho$ is a $Y$ eigenstate, which thus gives the
most stringent tests for these MDRs.
 
The precision and disturbance quantities for this example are
calculated from \erfand{precision}{disturbance}. They are
\bqa
\epsilon^2(Z) &=& \left<(U^\dagger (I \otimes Z) U - Z \otimes I)^2 \right> 
= 4 \left< I \otimes \ket{1}\bra{1} \right>, \nn \\ 
%
\eta^2(X) &=& \left< (U^\dagger (X \otimes I) U - X \otimes I)^2 \right> 
= 2 \left<  I \otimes (I - X) \right> . \nn
\eqa
For this particular measurement, both the precision and disturbance
quantities are independent of the input state: 
\begin{equation}
\epsilon(Z) = 2 |\sin \theta|\;,\;\; \eta(X) = \sqrt{2} |\cos \theta - \sin \theta|.
\end{equation}
The product in the Heisenberg-form MDR (\ref{falseMDR}) is thus 
\beq
H = \epsilon(Z)\eta(X)  = 2\sqrt{2}  |\sin \theta|   |\cos \theta - \sin \theta|.
\eeq
Now the pre-measurement uncertainties for an input $Y$ eigenstate are
\begin{equation}
\sigma(X) = 1\;,\;\; \sigma(Z) = 1.
\end{equation}
Thus the expression appearing in Ozawa's MDR is
\bqa
O &=& H + \epsilon(Z) \sigma(X) + \sigma(Z) \eta(X) \nn \\
&=& 2\sqrt{2}  |\sin \theta|   |\cos \theta - \sin \theta| + 2 |\sin \theta| + \sqrt{2} |\cos \theta - \sin \theta| \nn 
\eqa

\begin{figure}
\centerline{\includegraphics[width=12cm]{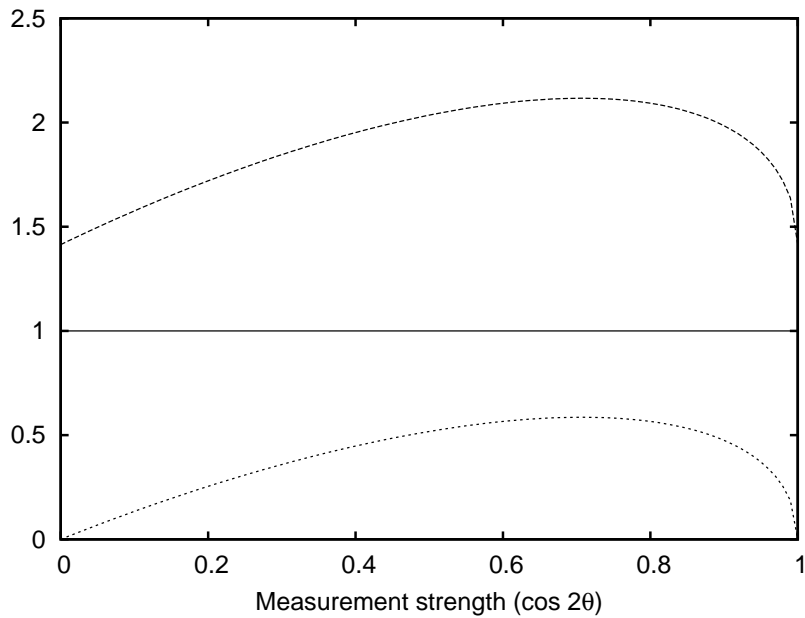}}
\caption{\label{graph} The quantities $H$ (dotted line) and $O$
(dashed line) appearing in the Heisenberg-form MDR and Ozawa's
universally valid MDR for the example qubit model. Both have the same
lower bound $C(X,Z)$ which equals $1$ here (solid line). In this case
the Heisenberg-form MDR is false for all measurement strengths.
}
\end{figure}

For both MDRs (\ref{falseMDR}) and (\ref{OzawaMDR}), the lower bound is
\beq
C(X,Z) = |\bra{\Psi} [X,Z] \ket{\Psi}|/2 = |\bra{\Psi} Y \ket{\Psi}| = 1.
\eeq
It is easy to verify that the Heisenberg-form MDR $H \geq C(X,Z)$ is
violated for all measurement strengths $\cos2\theta$. On the other
hand, Ozawa's universally valid MDR $O \geq C(X,Z)$ holds for all
$\theta$ as expected. See Fig.~\ref{graph}.

To measure Ozawa's $\epsilon(A)$ and $\eta(B)$ quantities it is
necessary in general to use a series of different weak measurements on
the initial signal, one for each eigenvalue of $A$ or $B$ [see
Eqs.~(\ref{wvpA}) and (\ref{wvpB}) respectively], as done for
$\eta(p)$ in Ref.~\cite{Mir07}. However, the sum of the weak-valued
probabilities will equal unity, and in the qubit case $X$ and $Z$ have
only two eigenvalues. Thus only a single weak measurement is needed,
so we can consider the observables $X$ or $Z$ themselves, and use, for
example,
\beq \label{sum2one}
2 \an{\Pi(x = \pm 1)_{\rm weak}} = 1 \pm \an{X_{\rm weak}}.
\eeq
Thus the initial weak measurement can be performed using a measurement
apparatus identical to that already defined above, based on a $CNOT$
gate, as shown in Fig.~\ref{qubit}. To avoid confusion with the {\em
meter} state, we write the input {\em probe} state as $\gamma \ket{0}
+\bar\gamma \ket{1}$. Here $\gamma , \bar\gamma \in \mathbb{R}_+$, so
that the measurement strength (which should be small) is $2\gamma^2 -
1$. The POVM elements \cite{NieChu00} corresponding to the two
outcomes $Z_p = \pm 1$ from reading out the probe are 
\beq \label{weak-meas}
E_\pm = \frac{1}{2} \left[1 \pm (2 \gamma^2 - 1) O \right],
\eeq
where $O=Z$ or $X$ as appropriate (see Fig.~\ref{qubit}). 

Now we will describe how the data from this experimental arrangement
is processed to yield $\epsilon(Z)$ and $\eta(X)$. First consider the
disturbance quantity $\eta(X)$. There are only two non-zero terms in
\erf{prec-delta}: 
\beq
\eta^2(X) = 4 P_{\rm wv}(\delta X = +2) + 4 P_{\rm wv}(\delta X = -2).
\eeq
Consider the $P_{\rm wv}(\delta X = +2)$ term. From \erf{using},  
$$
P_{\rm wv}(\delta X = +2) = P_{\rm wv} (X_i = -1 | X_f = 1) P(X_f = 1).
$$
The last factor equals the directly measurable probability from the
final signal read-out, which equals $1/2$ for the system in question.
The first factor is a {\em weak-valued} probability. From
\erf{sum2one}, it can be computed from directly measured joint
probabilities as follows:
\beq
2 P_{\rm wv} (X_i=\pm 1 | X_f) = 
 1 \pm \frac{\sum_{Z_p} Z_p P(Z_p | X_f)}{2 \gamma^2 - 1} .
\eeq
As above, $Z_p\in\{-1,1\}$ is the result of the read-out of the weak
probe, which effects the measurement as described by the POVM in
\erf{weak-meas} with $O=X$ in this instance. The precision quantity
$\epsilon(Z)$ can be obtained in exactly the same way, by changing the
probe interaction so that $O=Z$ in \erf{weak-meas}, and by replacing
the final read-out of the {\em signal} $X_f$ by a read-out of the {\em
meter} $Z_m$.

\begin{figure}
\centerline{\includegraphics[width=12cm]{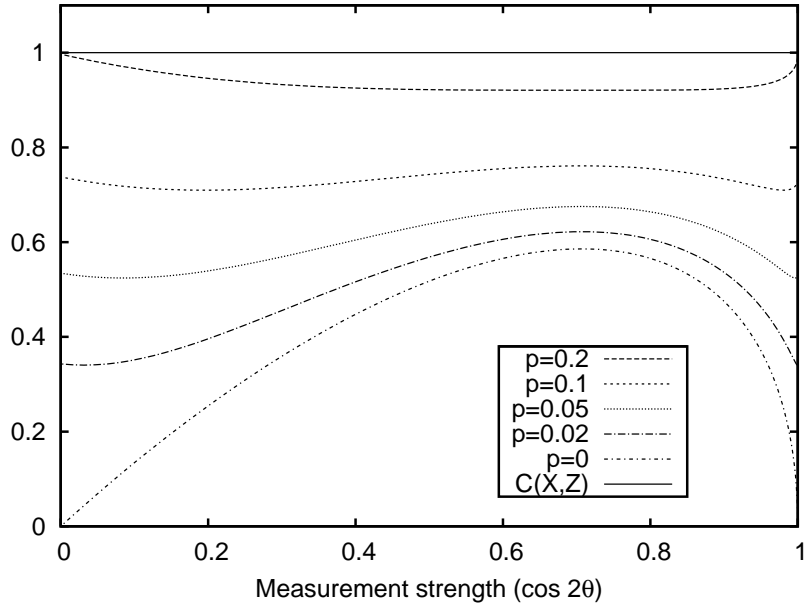}}
\caption{\label{simulation} The $H$ quantity from the Heisenberg-form
MDR evaluated numerically for the proposed experiment given in
Fig.~\ref{qubit} with each of the two $CNOT$ gates replaced by
non-ideal $CNOT$ operations parametrised by an error rate $p$ (see
text). A range of values of $p$ between $0$ and $0.2$ are shown (see
legend) with the values of $H$ tending to increase with increasing
$p$. The lower bound $C(X,Z)$ is the same as Fig.~\ref{graph} (solid
line). All of the curves shown here violate the lower bound of the
Heisenberg-form MDR.}
\end{figure}

\begin{figure}
\centerline{\includegraphics[width=12cm]{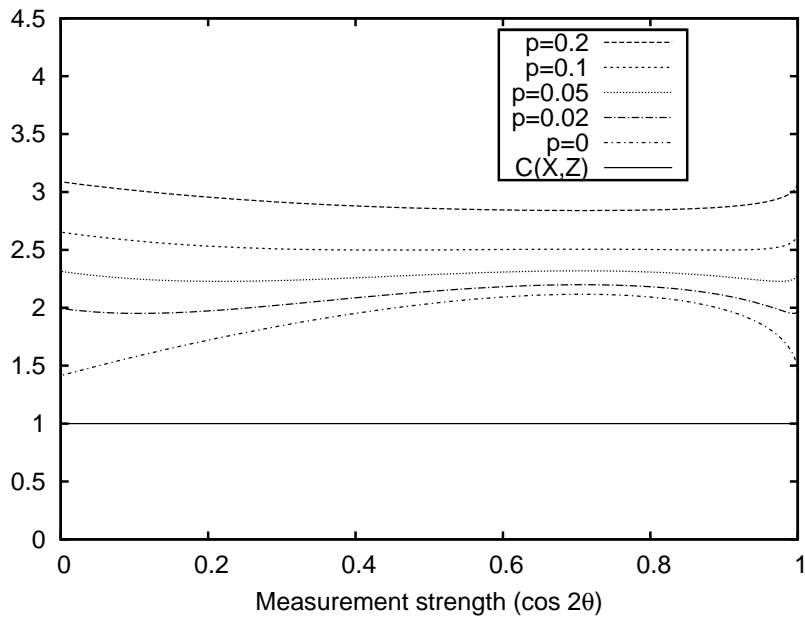}}
\caption{\label{simulation-ozawa} The $O$ quantity from Ozawa's MDR
plotted in the same manner as Fig.~\ref{simulation}. The values of $O$
tend to increase with increasing $p$ (see legend). All of the curves
shown here obey the lower bound of Ozawa's MDR.}
\end{figure}

To evaluate the feasibility of performing the proposed experiment, we
have performed a numerical simulation for imperfect $CNOT$ gates, with
results shown in Fig.~\ref{simulation} and
Fig.~\ref{simulation-ozawa}. The results in Fig.~\ref{simulation} plot
the $H$ quantity showing a violation of the Heisenberg-form MDR and
the results in Fig.~\ref{simulation-ozawa} plot the $O$ quantity
validating Ozawa's MDR. In this simulation the $CNOT$ gates are
replaced with non-ideal $CNOT$ operations. The non-ideal $CNOT$
consists of the ideal $CNOT$ gate with probability $1-p$, the identity
with probability $p/2$, and a swap gate \cite{NieChu00} with
probability $p/2$. This model is motivated by considering mode
mismatch in a standard experimental implementation of a $CNOT$ gate in
linear optics~\cite{Hof01,OBr03,Lang02}.

These simulations show that for the error model considered the
quantities of interest are independent of the probe measurement
strength. They also show that to demonstrate results similar to the
ideal (in particular, clearly showing a violation of the
Heisenberg-form MDR) requires error rates of 10\% or less for both
$CNOT$ gates. This should be feasible in a number of platforms for
quantum information processing.

\section{Conclusion}

In this paper we have proposed a method for experimentally determining
Ozawa's measurement precision $\epsilon(A)$ and disturbance $\eta(B)$
quantities, using weak measurements. This method requires no prior
information about how the initial system state is prepared, nor about
how the measurement apparatus under study operates (except for how its
outcomes encode a measurement of the relevant system observable $A$).
Moreover, this method is understandable classically and works for
classical as well as quantum systems.

Using this approach we suggest an example experiment using qubits
which would validate Ozawa's MDR whilst showing that the
Heisenberg-form MDR is violated. We have also performed numerical
simulations to quantify the operation of this three-qubit scheme
under an error model motivated by mode-mismatch in a linear optical
quantum computing realisation. Such an experiment, quantifying the
measurement--disturbance relation, would address a much-discussed
issue going back to the birth of modern quantum mechanics.

\ack

This work was funded by the ARC. HMW thanks Masanao Ozawa for
formative discussions.

\end{document}